# Dense Subgraph Maintenance under Streaming Edge Weight Updates for Real-time Story Identification

Albert Angel
University of Toronto
albert@cs.toronto.edu

Nikos Sarkas
University of Toronto
nsarkas@cs.toronto.edu

Nick Koudas
University of Toronto
koudas@cs.toronto.edu

Divesh Srivastava
AT&T Labs-Research
divesh@research.att.com

## ABSTRACT

Recent years have witnessed an unprecedented proliferation of social media. People around the globe author, every day, millions of blog posts, micro-blog posts, social network status updates, etc. This rich stream of information can be used to identify, on an ongoing basis, emerging stories, and events that capture popular attention. Stories can be identified via groups of tightly-coupled real-world entities, namely the people, locations, products, etc., that are involved in the story. The sheer scale, and rapid evolution of the data involved necessitate highly efficient techniques for identifying important stories at every point of time.

The main challenge in real-time story identification is the maintenance of dense subgraphs (corresponding to groups of tightly-coupled entities) under streaming edge weight updates (resulting from a stream of user-generated content). This is the first work to study the efficient maintenance of dense subgraphs under such streaming edge weight updates. For a wide range of definitions of density, we derive theoretical results regarding the magnitude of change that a single edge weight update can cause. Based on these, we propose a novel algorithm, DYNDENS, which outperforms adaptations of existing techniques to this setting, and yields meaningful results. Our approach is validated by a thorough experimental evaluation on large-scale real and synthetic datasets.

## 1. INTRODUCTION

Recent years have witnessed an unprecedented proliferation of social media. Millions of people around the globe author on a daily basis millions of blog posts, micro-blog posts and social network status updates. This content offers an uncensored window into current events, and emerging stories capturing popular attention.

For instance, consider the U.S. military strike in Abbottabad, Pakistan in early May 2011, which resulted in the death of Osama bin Laden. This event was extensively covered on Twitter, the popular micro-blogging service, significantly in advance of traditional media, starting with the live coverage of the operation by an (unwitting) local witness, to millions of tweets around the world providing a multifaceted commentary on every aspect of the story. Similar, if fewer, online discussions cover important events on an everyday basis, from politics and sports, to the economy and culture (notable examples from recent years range from the death of Michael Jackson, to revolutions in the Middle East and the economic recession). In all cases, stories have a strong temporal component, making timeliness a prime concern in their identification.

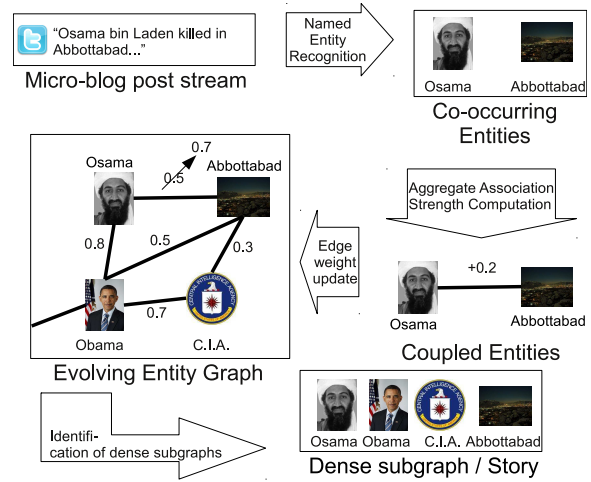

**Figure 1: Real-time identification of "bin Laden raid" story, and connection to** ENGAGEMENT

Interestingly, such stories can be identified by leveraging the real-world entities involved in them (e.g. people, politicians, products and locations) [26]. The key observation is that each post on the story will tend to mention the same set of entities, around which the story is centered. In particular, as post length restrictions or conventions typically limit the number of entities mentioned in a single post, each post will tend to mention entities corresponding to a single facet of a story. Thus, by identifying pairs of entities that are strongly associated (recurrently mentioned together), one can implicitly detect facets of the underlying event of which they are the main actors. By piecing together these aspects, the overall event of interest can be inferred.

For example, in the case of the U.S. military strike mentioned above, one facet, consisting of people discussing the raid, is centered around "Abbottabad" where the raid took place, and the involvement of the "C.I.A."; another thread commenting on the presidential announcement, involves "Barack Obama" and "Osama bin

Laden"; and so on. The resulting overall story at some point of time involves the union of these entities. Such sets of entities can be then used by users of systems such as Grapevine [3] to enable the interactive exploration of the story.

Given a measure to quantify the strength of association between two entities (such as the Log-likelihood ratio [26], the $\chi^2$ measure, or the correlation-coefficient [5], etc.), one can abstract the real-time stream of posts giving rise to an evolving (weighted) entity graph, denoting the pairwise entity association strength[1]. An important story can then be identified via a cohesive group of strongly associated entity pairs; i.e. a dense subgraph in the entity graph, given an appropriate definition of density. Moreover, note that, as the entities in a story need to be presented to users to facilitate navigation, story cardinality needs to be constrained to moderate sizes; after all, it would not be very interesting or helpful to present users with a story centered around 100 main entities. This process is illustrated in Figure 1.

Every post that is published, results in the weight update of one or more edges in the entity graph. The high frequency of post generation, coupled with our need for timely reporting of emerging stories, necessitates that the identification of dense structures in the entity graph be highly efficient. This work thus addresses the problem of dENse subGrAph maintenance for edGE-weight update streaMs under sizE constraiNTs, or ENGAGEMENT for brevity. Besides being useful as-is for identifying stories from social media in real-time, solutions to this problem can also be used as building blocks for more complex computations; e.g. identified dense subgraphs can undergo diversification before being presented to the user [2], or they can be reranked taking their external sparsity into account, in order to identify (soft) clusters of associated entities.

Addressing ENGAGEMENT at web scales presents several challenges. Principal among these is that, a change in the weight of a single edge, can impact the density of many subgraphs, necessitating a potentially unbounded exploration of the entity graph. Thus, any efficient solution to ENGAGEMENT needs to incrementally maintain dense subgraphs, without recomputing them from scratch. Moreover, there does not exist a single definition of graph density suitable for all scenarios; selecting the most appropriate definition for a given setting depends, for instance, on the perceived relative importance of having large, versus well-connected, dense subgraphs. Thus, solutions to ENGAGEMENT need to be applicable under general notions of density; however, existing techniques are only applicable to limited subsets of this problem.

In this context, in this work we propose DYNDENS, an efficient algorithm for ENGAGEMENT. We theoretically quantify the magnitude of change in dense subgraphs that a single edge weight update can cause. Based on this, we show how maintaining some sparse subgraphs, in addition to dense ones, enables the incremental maintenance of dense subgraphs. The resulting algorithm, DYNDENS, makes use of an efficient index for subgraphs, which decreases memory consumption and processing effort. It is complemented by theoretically sound heuristics, that can offer improved performance. A comprehensive experimental evaluation on real and synthetic data highlights the effectiveness of our approach.

To summarize, our main contributions in this work are:

i) Motivated by the need to identify emerging stories in real-time, for a wide range of measures of entity association, we formalize the problem of dENse subGrAph maintenance for edGE-weight update streaMs under sizE constraiNTs (ENGAGEMENT), for a very broad notion of graph density.

ii) We propose an efficient algorithm DYNDENS, based on a novel quantification of the maximum possible change caused by a single edge weight update. By maintaining a small number of sparse subgraphs, DYNDENS is able to efficiently and incrementally compute dense subgraphs.

iii) We design an efficient dense subgraph index, which decreases memory consumption and processing effort, and propose theoretically sound heuristics for DYNDENS that can offer improved performance.

iv) We validate our techniques via a thorough experimental evaluation on both real and synthetic datasets.

The remainder of this paper is organized as follows: After providing a formal problem statement in Section 2, we present our proposed algorithm DYNDENS in Section 3. We explore the theoretical basis for DYNDENS in Section 4, evaluate the proposed techniques in Section 5, and discuss some improvements to DYNDENS in Section 6. Finally, we review related work in Section 7, and conclude in Section 8.

## 2. FORMALIZATION

Let us now turn to defining ENGAGEMENT. At a high level, let us consider a weighted graph, with a constant number of vertices. At every discrete time interval, the weights of one or more edges are adjusted (including potentially edge additions and removals). The goal is to maintain, at each point of time, all subgraphs with "density" greater than a given threshold.

**Connections to real-time story identification:** Before fully formalizing the problem, let us first draw some connections to its application in real-time story identification. In this context, vertices correspond to real-world entities, and edge weights to their (current) pairwise association strengths (the choice of association strength measure will depend on characteristics of the specific problem instance; in Section 5 we discuss several such choices). We assume that a procedure exists for processing streams of (entity-annotated[2]) posts, and generating the appropriate edge weight updates at each time interval (in Section 5 we discuss such procedures for a variety of measures of interest).

**Data model:** We represent the problem domain as i) a complete weighted graph $G = (V, E)$ with $N$ vertices, where $w_{ij}$ is the weight of edge between nodes $i$ and $j$; and ii) a stream of edge weight updates of the form $update_i = (a, b, \delta)$, signifying that at time instant $i$, the weight of the edge between vertices $a$ and $b$ changed from $w_{ab}$ to $w_{ab} + \delta$.

**Density:** We define subgraph density as follows: for every subgraph $C \subseteq V$, its density is $dens(C) = \frac{score(C)}{S_{|C|}}$, where $score(C) = \sum_{i,j \in C \land i<j}(w_{ij})$. $S_n$ is a function quantifying the relative importance of a subgraph's cardinality, $n$, to its density; with the appropriate choice of $S_n$, virtually all quantifications of graph density can be represented.

Note that we do not consider counter-intuitive quantifications of graph density, such as (but not limited to) a definition of density where the removal of a vertex from an unweighted clique results in an increase of its density. To safeguard against such quantifications of density, we require that $S_n$ have the following intuitive monotonicity properties: $\frac{n}{n-1} \leq \frac{S_n}{S_{n-1}} \leq \frac{n}{n-2}$.[3] This encompasses the full spectrum of choices of density functions commonly used in the literature; typical choices include $S_n = \frac{n \cdot (n-1)}{2}$ (thus density is defined as the average edge weight, favoring small, dense subgraphs; we term this instantiation AVGWEIGHT), and $S_n = n$ (thus density represents a generalized average node "degree", favoring large subgraphs; we term this case AVGDEGREE).

---
[1]The association measure can also incorporate notions of recency of association, e.g. by including some form of temporal decay.

[2]The precise procedure used for identifying named entities in documents, e.g. [3], is orthogonal to this work.

[3]Observe that if $\frac{S_n}{S_{n-1}} > \frac{n}{n-2}$, the density of an unweighted clique will increase if vertices are removed. Moreover, observe that if $\frac{n}{n-1} > \frac{S_n}{S_{n-1}}$, in an unweighted graph the density of an 3-vertex clique $K_3$ will increase if it is augmented by a single vertex, connected with a single edge to one of the clique vertices.

**Cardinality constraint:** Finally, let $N_{max}$ be a (user-specified) maximum cardinality for subgraphs of interest. (In the context of real-time story identification, this constraint ensures that any subgraphs identified are small enough to be used for navigation / exploration purposes - cf. Section 1).

ENGAGEMENT: Given the above, the goal of ENGAGEMENT is to maintain, at every point of time $i$, the subgraphs (vertex subsets) with density over a given threshold $T$, subject to cardinality constraints, i.e. $\{V_j | V_j \subseteq V \land dens(V_j) \geq T \land |V_j| \leq N_{max}\}$. We term these output-dense subgraphs.

**Notation:** Before going into the details of our proposed approach, let us introduce some useful notation.

We denote each vertex by a natural number, so $V = \{1, \cdots, N\}$ denotes the set of vertices in $G$.

Let $\hat{e}_i$ be the $i$'th basis vector (an $N$-dimensional vector, with value 1 in its $i$'th coordinate, and 0 elsewhere). We will denote a subset $C \subseteq V$ by its corresponding vector $\vec{c} = \sum_{i \in C} \hat{e}_i$, and will sometimes refer to either interchangeably; we will also on occasion denote the cardinality of subset $C$ as $|\vec{c}|$.

Let $\vec{\Gamma}_u$ be the neighborhood vector of vertex $u$: $\vec{\Gamma}_u = (w_{1u}, w_{2u}, \cdots, w_{Nu})$.

For convenience, we will also make use of the following normalized version of $S_n$: Let $g_n = \frac{S_n}{n \cdot (n-1)}$. By the monotonicity properties of $S_n$, it follows that $g_n \leq g_{n-1}$.

Unless explicitly stated, we will focus on the time instant where the weight of the edge between vertices $a$ and $b$ is updated from $w_{ab} = w$ to $w + \delta$. Whenever a quantity $X$ can be affected by this update, we will denote its value before the update as $X^-$ and its value after the update as $X^+$. We omit this superscript when it does not affect results in any way. For example, $w_{ab}^- = w$, $w_{ab}^+ = w + \delta$.

## 3. THE DYNDENS APPROACH

Let us now discuss how our proposed algorithm, DYNDENS, identifies, at every point of time, all output-dense subgraphs.

**Dense subgraphs and growth property:** Observe that there is an inherent tradeoff in the set of subgraphs that DYNDENS will maintain, which we term "dense" subgraphs. At one extreme, DYNDENS could opt to maintain only output-dense subgraphs, with the other extreme being to maintain *all* subgraphs. However, neither of these is desirable: the former because it does not enable incremental computation of output-dense subgraphs, the latter due to its prohibitive costs. We will subsequently (Section 4.2) formally quantify this tradeoff. For now, loosely speaking, we will say that $C$ is a dense subgraph iff it has density greater than a given threshold $T_{|C|}$ (which is a function of the cardinality of $C$), and cardinality of at most $N_{max}$ (for a complete list of density-related terms used in this work cf. Table 1). $T_n$ is defined in a manner that ensures that every dense graph with $n$ vertices has at least one dense subgraph with $n-1$ vertices (thus it is possible to identify all dense subgraphs by "growing" dense subgraphs of smaller cardinalities).

Specifically, $T_n$ is a monotonically increasing function of $n$ with the property $T_n \cdot g_n > T_{n-1} \cdot g_{n-1}$. At a high level, this monotonicity property ensures the desired containment property mentioned earlier (see Section 4 for details[4]). Moreover, we require that $T_{N_{max}} = T$.[5] We discuss the concrete instantiation of $T_n$ used by DYNDENS in Section 4.2.

[4] Another way to view dense graphs is the following: Consider the

**Table 1: Definitions of density-related properties**

| Subgraph $C$ is $\cdots$ | iff |
|---|---|
| Static properties | |
| dense | $dens(C) \geq T_{|C|}$ |
| sparse | $dens(C) < T_{|C|}$ |
| output-dense | $dens(C) \geq T$ |
| too-dense | $dens(C) \geq T_{|C|+1}$ |
| Dynamic properties | |
| stable-dense | $dens(C)^- \geq T_{|C|} \land dens(C)^+ \geq T_{|C|}$ |
| newly-dense | $dens(C)^- < T_{|C|} \land dens(C)^+ \geq T_{|C|}$ |
| losing-dense | $dens(C)^- \geq T_{|C|} \land dens(C)^+ < T_{|C|}$ |

**Table 2: Summary of main symbols used**

| Symbol | Description |
|---|---|
| $V$ | Set of vertices in graph |
| $N$ | Number of vertices in graph |
| $w_{ij}$ | Weight of edge between vertices $i$ and $j$ |
| $\vec{\Gamma}_u$ | Neighborhood vector of vertex $u$ |
| $dens(C)$ | Density of $C$ |
| | $dens(C) = \frac{\sum_{i,j \in C \land i < j}(w_{ij})}{S_{|C|}}$ |
| $S_n$ | Quantifies relative importance of subgraph cardinality $n$ to density |
| $g_n$ | Normalized version of $S_n$: $g_n = \frac{S_n}{n \cdot (n-1)}$ |
| AVGWEIGHT | Case where $S_n = n(n-1)/2$ |
| SQRTDENS | Case where $S_n = \sqrt{n}(n-1)$ |
| AVGDEGREE | Case where $S_n = n$ |
| $N_{max}$ | Max. cardinality of subgraph to be returned |
| $T$ | Min. density for a subgraph to be returned |
| $T_n$ | Min. density for subgraph of cardinality $n$ to be dense |
| $\delta_{it}$ | Tunable parameter of DYNDENS, influences $T_n$ |
| $a, b$ | Vertices that were just updated |
| $x^-$ | quantity x before the update |
| $x^+$ | quantity x after the update |
| $w$ | Weight of edge $(a,b)$ before the update, ie. $w_{ab}^-$ |
| $w + \delta$ | Weight of edge $(a,b)$ after the update, ie. $w_{ab}^+$ |

**Edge weight updates:** The basic operation of DYNDENS is to maintain dense subgraphs, following the update of the weight of an edge $(a, b)$, from $w$ to $w + \delta$. If this impacts the set of output-dense subgraphs, the latter is updated as well. Handling updates with $\delta < 0$ (i.e. where the weight of an edge decreases) is straightforward: all dense subgraphs containing both $a$ and $b$ are examined, and their density is decreased by an appropriate amount. If they are no longer output-dense, this is reported; if, in addition, they are no longer dense (losing-dense), they are evicted from the index.

**Positive updates:** Of greater interest is the case where $\delta > 0$, i.e. the edge weight update corresponds to an increase in weight. In this case, additional subgraphs, that were not dense prior to the update, might now be dense (newly-dense subgraphs). DYNDENS leverages the growth property to compute these as follows:

---

measure $normDens(C) = \frac{dens(C)}{T_{|C|}}$, consisting of a density measure, normalized by the threshold function $T_n$; a graph $C$ is dense iff it has $normDens(C) \geq 1$. While $normDens(C)$ is not a suitable measure of density per se, it has the following important growth property: every graph $C$ has a subgraph $C'$ of cardinality $|C'| = |C| - 1$ with $normDens(C') \geq normDens(C)$. This containment/growth property additionally implies that, if there are no dense subgraphs of cardinality $n$, there can be no dense subgraphs of any cardinality $> n$.

[5] Recall that $T_n$ is an increasing function of $n$, and the set of maintained subgraphs needs to include all output-dense subgraphs of cardinality $\leq N_{max}$ having density $\geq T$.

**Algorithm 1** Algorithm DYNDENS
**Input:** Updated edge $(a, b)$, magnitude of update $\delta$
1: **if** $\delta < 0$ **then**
2:    Update the density of all dense subgraphs containing $a$ and $b$; evict losing-dense subgraphs from the index; report any subgraphs that are no longer output-dense
3:    **return**
4: **for all** dense subgraphs $C$ st. $a \in C \vee b \in C$ **do** {// including $C = \{a, b\}$ if it is newly-dense}
5:    **if** $a \notin C$ **or** $b \notin C$ **then**
6:      **if** $C$ should be cheap-explored and $C \cup \{a, b\}$ is newly-dense **then**
7:         Add $C \cup \{a, b\}$ to the index, report it if it is output-dense
8:         **explore**$(C \cup \{a, b\}, 2)$
9:    **else**
10:      Update the density of $C$, report it if it just became output-dense
11:      **explore**$(C, 1)$

**Algorithm 2** Procedure **explore**$(C, i)$
**Input:** Subgraph $C$. Iteration number $i$
1: **if** $C$ was not too-dense before the update and $i \leq \lceil \frac{\delta}{\delta_{it}} \rceil$ and $|C| < N_{max}$ **then**
2:    **if** $C$ is too-dense **then**
3:      **for all** $y \notin C$ **do** {// Explore-All}
4:         Add $C \cup \{y\}$ to the index; report it if it is output-dense
5:         **explore**$(C \cup \{y\}, i + 1)$
6:    **else**
7:      **for all** neighbors $y$ of $C$ **do**
8:         **if** $C \cup \{y\}$ is newly-dense **then**
9:            Add $C \cup \{y\}$ to the index; report it if it is output-dense
10:            **explore**$(C \cup \{y\}, i + 1)$

**Cheap explore**: DYNDENS will try to augment all dense subgraphs containing either $a$ or $b$, with $b$ or $a$, respectively; resulting newly-dense subgraphs will be inserted into the dense subgraph index. In some cases, this step alone is sufficient and/or can be applied only to a subset of these subgraphs (cf. Section 6) for details).

**Explore**: DYNDENS will try to augment dense subgraphs containing both $a$ and $b$, with one neighboring vertex; resulting newly-dense subgraphs will be inserted into the dense subgraph index.

**Exploration iterations**: The above procedure may need to be performed iteratively for newly-dense subgraphs discovered via exploration or cheap exploration. Interestingly, the iteration depth is upper bounded by a corollary of the growth property. Specifically, in Section 4.2, we define $T_n$ parametrized by a parameter $\delta_{it}$ that indirectly controls the number of dense subgraphs maintained by DYNDENS. As we show in Section 4, we can guarantee that at most $\lceil \frac{\delta}{\delta_{it}} \rceil$ iterative exploration iterations need to be performed, in order to identify all newly-dense subgraphs, following an edge weight update of magnitude $\delta$.

**Explore all**: In a few cases, the above exploration may need to be performed on non-neighboring nodes as well, resulting in a very costly procedure. In most cases, DYNDENS avoids performing this procedure via a better, implicit representation of some dense subgraphs in the index (cf. Section 3.2.3).

In one sentence, DYNDENS explores the neighborhood of some materialized dense subgraphs, using pruning conditions for when to stop exploring around a subgraph. The remainder of this section aims to fill in the blanks in the preceding sentence. We discuss the workings of DYNDENS, and illustrate them with a practical example in Section 3.1, followed by important technical details in Section 3.2. We defer the exposition of the theoretical results on which DYNDENS is based till Section 4.

## 3.1 The DYNDENS Algorithm

Let us now discuss DYNDENS in greater detail, with reference to Algorithm 1. At a high level, DYNDENS maintains an in-memory index of all dense subgraphs (we defer discussing index implementation details to Section 3.2); at every edge weight update, it outputs information regarding subgraphs that became, or stopped being output-dense. If the edge weight update was negative, only some index maintenance needs to be done (line 2). Otherwise, some stable-dense subgraphs containing $a$ and/or $b$ are further ex-amined (lines 4-11). Note that, to ensure correctness, also the subgraph $\{a, b\}$ may be examined, even if it was not present in the index (base case in line 4). Subgraphs in the index containing only one of $a, b$ are cheap-explored, if needed[6] (line 6).

Subgraphs in the index that contain both $a$ and $b$, as well as newly-dense subgraphs previously identified, are subsequently explored (line 11) - i.e. DYNDENS will try to augment them with a neighboring node (we defer discussing the precise details on how this is done efficiently to Section 3.2). This will be recursively repeated on any newly-dense subgraphs discovered up to $\lceil \frac{\delta}{\delta_{it}} \rceil$ times (the theoretical results that enable this bounding are discussed in Section 4). A high-level description of the exploration procedure is shown in Algorithm 2.

Algorithm 2 will first ensure that the subgraph should be explored. Specifically, the subgraph should not have been too-dense before the update (line 1), for otherwise its dense supergraphs would have been stable-dense, and hence already identified. Moreover, as previously mentioned, DYNDENS will not explore around any subgraph more times than necessary. Finally, in a few cases, explored subgraphs will need to be augmented with every other vertex, not just neighboring ones (Explore-All; line 2). As the latter is a costly procedure, in Section 3.2.3 we will present a way to mitigate the associated cost.

*Execution example.* To illustrate the workings of DYNDENS, let us examine a simple example of its execution. Consider the sample entity graph of Figure 2(a), and assume an AVGWEIGHT definition of density (i.e. the density of a subgraph is its average edge weight), a density threshold of $T = 1$, and a maximum desired subgraph cardinality of $N_{max} = 4$. Assume that $\delta_{it}$ has been set to 0.15, so that the thresholds $T_n$, for subgraphs of cardinality $n$ to be considered dense are $T_2 = 0.9, T_3 = 0.975$ and $T_4 = T = 1$ (cf. Section 4.2 for details). Thus, the dense subgraphs for this graph are shown in Figure 2(b) (output-dense subgraphs are emphasized). Finally, assume that the weight of edge $(1, 2)$ is updated from 0.8 to 0.95 ($\delta = \delta_{it} = 0.15$). Let us examine how DYNDENS will handle this update; to facilitate this discourse, the newly-dense subgraphs that are inserted into the index are shown in the bottom half of Figure 2(b).

At a high level, DYNDENS will examine $\{1, 2\}$, as well as all dense subgraphs containing vertex 1 and/or 2 (Algorithm 1, line 4), i.e. $\{1, 3\}, \{1, 4\}, \{2, 3\}, \{2, 4\}, \{1, 3, 4\}, \{2, 3, 4\}$. $\{1, 2\}$ will be added to the index (Algorithm 1, line 10), and will be explored (line 11). Its exploration will entail the addition of newly-dense subgraphs $\{1,2,3\}$ and $\{1,2,4\}$ to the index (Algorithm 2, line 8); the former will also be reported as output-dense. Since $\frac{\delta}{\delta_{it}} = 1$, these newly-dense subgraphs will not be further explored (Algorithm 2, line 10 and line 1). Moreover, during this exploration subgraph $\{1,2,5\}$ will be examined, but as its density is less than $T_3$, it will not be added to the index.

---

[6]For instance, subgraphs that were too-dense need not be explored, as, by definition, their dense supergraphs would have been stable-dense, and hence already identified. Moreover, this step can also be skipped in other circumstances, cf. Section 6 for details.

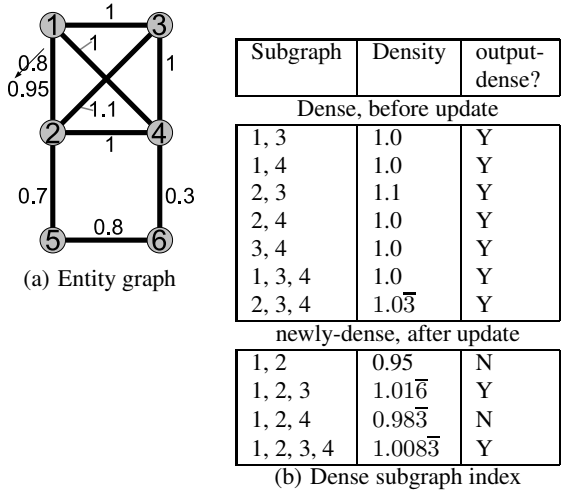

(a) Entity graph

(b) Dense subgraph index

| Subgraph | Density | output-dense? |
|---|---|---|
| Dense, before update | | |
| 1, 3 | 1.0 | Y |
| 1, 4 | 1.0 | Y |
| 2, 3 | 1.1 | Y |
| 2, 4 | 1.0 | Y |
| 3, 4 | 1.0 | Y |
| 1, 3, 4 | 1.0 | Y |
| 2, 3, 4 | $1.0\overline{3}$ | Y |
| newly-dense, after update | | |
| 1, 2 | 0.95 | N |
| 1, 2, 3 | $1.01\overline{6}$ | Y |
| 1, 2, 4 | $0.98\overline{3}$ | N |
| 1, 2, 3, 4 | $1.008\overline{3}$ | Y |

**Figure 2: Execution example**

DYNDENS will also cheap-explore subgraphs $\{1,3\}$, $\{1,4\}$, $\{2,3\}$, $\{2,4\}$ (Algorithm 1, line 6). This will result in subgraphs $\{1,2,3\}$, $\{1,2,4\}$ being examined (twice) (Algorithm 1, line 7); as they are already present in the index, this will not affect anything. Moreover, DYNDENS will attempt to explore these subgraphs (Algorithm 1, line 8); however, since $\frac{\delta}{\delta_{it}} = 1$, they will not be explored (Algorithm 2, line 1).

Finally, DYNDENS will cheap-explore subgraphs $\{1,3,4\}$ and $\{2,3,4\}$. The first cheap exploration will result in newly-dense subgraph $\{1,2,3,4\}$ being added to the index, and reported as output-dense (Algorithm 1, line 7); the second one will revisit this subgraph, and do nothing. Moreover, in both cases, since $|\{1,2,3,4\}| = 4 \geq N_{max}$, these subgraphs will not be explored (Algorithm 2, line 1).

**Observation:** From the simplified execution example presented above, one can observe that DYNDENS (as currently presented) can end up performing redundant computations; e.g. some subgraphs are examined unnecessarily many times. Subsequently, in Section 3.2.2 and Section 6, we discuss how to reduce such unnecessary computations.

## 3.2 Implementation Considerations

Having presented DYNDENS at a high level, let us now see some important considerations that arise when implementing it in practice. We first introduce the underlying indexing structure used by DYNDENS in Section 3.2.1; this index also enables DYNDENS to avoid redundant computations (Section 3.2.2) as well as the costly operation of explore-all (Algorithm 2, line 2 cf. Section 3.2.3).

### 3.2.1 Index

DYNDENS requires an efficient index for both the evolving graph itself, as well as for dense subgraphs. For the graph index, maintaining node adjacency lists is sufficient (i.e. a mapping $\forall u \in V : u \to \vec{\Gamma}_u$); this also enables the efficient exploration of a subgraph

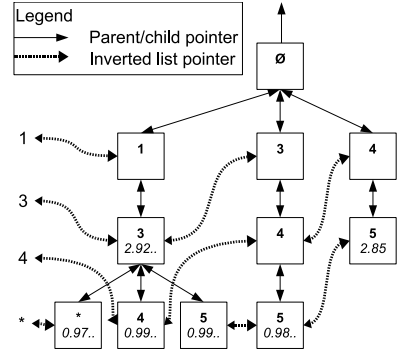

**Figure 3: Dense subgraph index**

(via merging the relevant adjacency lists[7]).

The dense subgraph index is more interesting to examine, as it needs to efficiently support several functionalities. To name a few: for every dense subgraph, access to its vertices, cardinality and density; insertion, update and deletion of dense subgraphs from the index; iteration over all dense subgraphs containing vertices $a$ or $b$, where each subgraph must be accessed exactly one time (needed for positive edge weight updates); and for a given dense subgraph $C$, and a given vertex $u$, access to subgraph $C \cup \{u\}$, and insertion of $C \cup \{u\}$ into the index if it is not already present (needed for exploration). Moreover, as DYNDENS needs to perform frequent random accesses on dense subgraphs, the index needs to be in-memory, so maintaining a low memory footprint is important. As most dense subgraphs will tend to have high overlap, the dense subgraph index should minimize the amount of redundant information stored.

To address these requirements posed by DYNDENS, we propose the following in-memory index. Each subgraph has a unique id corresponding to its location in memory; it is also represented by its (sorted) set of vertices. DYNDENS will maintain a prefix tree of dense subgraphs, illustrated in Figure 3. Each node in the prefix tree contains pointers to its children, indexed by vertex id, a pointer to its parent, as well as information (such as cardinality and density) on the dense subgraph it represents, if applicable. Figure 3 shows a view of the index when subgraphs $\{1,3\}, \{1,3,4\}, \{1,3,5\}, \{3,4,5\}, \{4,5\}$ are dense (ignore node labeled $*$ for now), along with the density of each subgraph.

Additionally, to enable effective iteration over dense subgraphs containing one or two given vertices, DYNDENS will also maintain inverted lists, i.e. a mapping from vertices to (pointers to) all subgraphs containing a vertex. To decrease the size of inverted lists, the inverted list of a vertex $u$ will only contain tree nodes where the lexicographically largest vertex is $u$. Thus, in order to iterate over all subgraphs containing $u$, DYNDENS will iterate over all subgraphs in its inverted list, and their tree descendants. Furthermore, to facilitate inverted list maintenance, inverted lists are implemented as linked lists of prefix tree nodes (shown in Figure 3 as dashed arrows). Inverted lists are updated whenever a new node is created, or when a leaf node is deleted. Moreover, if the deletion of a leaf node results in its parent having no children, and representing no dense subgraph, the parent will be recursively deleted.

Our proposed dense subgraph index efficiently addresses the requirements of DYNDENS. Specifically, the prefix tree enables DYNDENS to reduce its memory footprint, by not storing redundantly many overlapping dense subgraphs. Moreover, looking up $C \cup \{u\}$ is $O(|C|+1)$ in all cases (and $O(1)$ if vertex $u$ is lexicographically greater than any other vertex in $C$); after a look-up, update or insertion into the index is $O(1)$. Enumerating the vertices in a subgraph $C$ is $O(|C|)$, via parent pointer traversal. Deleting a subgraph $C$ from the index is $O($ number of leaf nodes deleted$)$; this is typically $O(1)$ and at worst $O(|C|)$, due to the design of the prefix tree with embedded inverted lists.

---

[7]Specifically, when exploring subgraph $C$, DYNDENS will compute $\vec{\Gamma}_C = \sum_{v \in C} \vec{\Gamma}_v$; for every vertex $u \notin C$, the score of $C \cup \{u\}$ can be computed as $score(C \cup \{u\}) = score(C) + \vec{\Gamma}_C \cdot \hat{e}_u$.

### 3.2.2 Avoiding redundant computation

Besides efficiently providing the requisite functionality for DYN-DENS, our proposed dense subgraph index can also be used (i) to ensure that subgraphs that were dense before the update are examined exactly once (required for the correctness of DYNDENS), and (ii) to greatly reduce the number of newly-dense subgraphs examined more than once (without sacrificing correctness).

The former (i) can be guaranteed by fixing the order in which dense subgraphs are examined. Specifically, if subgraphs containing vertices $a$ and/or $b$ need to be examined, and assuming $a < b$ (lexicographically), DYNDENS will traverse the subtrees of all index nodes on the inverted list corresponding to $b$. Subsequently, it will traverse the subtrees of index nodes on the inverted list corresponding to $a$, stopping the traversal whenever a $b$ node is encountered. This procedure is aided by flags that are set on a per-index node basis, to help DYNDENS distinguish newly-dense subgraphs in the index.

For the latter (ii), we leverage the theoretical result that all newly-dense subgraphs can be identified in at most $\lceil \frac{\delta}{\delta_{it}} \rceil$ exploration iterations (Section 4). Upon insertion into the index, dense subgraphs are annotated with the exploration iteration at which they were identified ($i$ in Algorithm 2); these annotations persist until the end of Algorithm 1. Algorithm 2 will operate as above for subgraphs not annotated with an iteration number, or annotated with an iteration number greater than the current $i$. Otherwise, the subgraph does not need to be further examined.

### 3.2.3 Implicit representation of too-dense subgraphs

Having introduced the dense subgraph index used by DYNDENS, let us revisit a challenge posed by the presence of too-dense subgraphs, and show how the index can be leveraged to overcome it.

Recall that a subgraph is too-dense iff, after adding any other vertex to it, it is still dense. Thus, when exploring a too-dense subgraph, DYNDENS needs to consider its cartesian product with the entire set of vertices $V$, resulting in $|V|$ dense subgraph insertions into the index (explore-all, Algorithm 2, line 2). This is a very costly procedure; unsurprisingly, it was experimentally found to dominate all other processing costs, in cases where too-dense subgraphs existed (cf. Section 5.1).

To avoid this cost, we propose a modification to the dense subgraph index, which we term IMPLICITTOODENSE. At a high level, it entails the implicit representation of supergraphs of too-dense subgraphs, so that explore-all will only examine/insert into the index a small number of dense subgraphs.

Specifically, we introduce a fictitious vertex named $*$, which is lexicographically larger than all other vertices. For every too-dense subgraph $C$, the index will store a subgraph $C \cup \{*\}$, representing all $C \cup \{y\}$ where $y$ is a vertex disconnected from $C$; these supergraphs of $C$ will not be explicitly inserted in the index. Given this convention, DYNDENS will handle the explore-all procedure of a subgraph $C$ that just became too-dense by normally exploring all neighbors of $C$ (as in Algorithm 2, line 7), and inserting the subgraph $C \cup \{*\}$ into the index. For instance, revisiting Figure 3, assume subgraph $\{1,3\}$ is too-dense. Rather than exploring, and inserting into the index all its disconnected supergraphs $\{1,3,6\}, \{1,3,7\}, \cdots, \{1,3,|V|\}$, DYNDENS has only inserted a node representing $\{1,3,*\}$.

In the unlikely event $C \cup \{*\}$ needs to be explored at any time (corresponding to the exploration of all supergraphs of $C$ augmented with one disconnected vertex), DYNDENS will try instead to augment $C$ with all edges in the graph that are not incident on $C$.

Because every vertex $a$ is potentially contained in $C \cup \{*\}$, whenever an iteration is performed on the index (Algorithm 1, line 4), the inverted list corresponding to $*$ needs to be examined as well. This inverted list also needs to be maintained during negative edge weight updates, if a subgraph stops being too-dense. Finally, note that whenever dealing with a subgraph represented by a $*$ index entry, DYNDENS also needs to ensure that the subgraph is not explicitly represented elsewhere in the index, which is, however, a very efficient operation.

As we verify experimentally (Section 5.1), the above IMPLICIT-TOODENSE modification to the index offers significant performance benefits to DYNDENS.

## 4. THEORETICAL RESULTS

Having introduced our proposed DYNDENS algorithm, in this section we elaborate on its theoretical underpinnings. We first prove its correctness, by deriving a bound on the number of exploration iterations that are required, as a function of the magnitude of the edge weight update performed (this is the basis of DYNDENS, cf. Algorithm 2, line 1). Specifically, Section 4.1 presents a general result, on when a single exploration iteration per stable-dense subgraph is sufficient. Section 4.2 provides a concrete instantiation for $T_n$ (recall that $T_n$ determines the relationship between dense and output-dense subgraphs), based on which the desired bound is then obtained in Section 4.3. Due to space constraints, detailed proofs, and results pertaining to the complexity of DYNDENS are omitted; these can be found in [4].

**Formalization :** The notion of exploration iterations performed by DYNDENS has been used throughout its description; before presenting theoretical results on them, this would be a good opportunity to formalize this notion.

Let $C_A = \{C \cup \{b\} | C \subseteq V \wedge a \in C \wedge b \notin C \wedge C \text{ is stable-dense}\}$ be the set of graphs consisting of a stable-dense subgraph containing $a$, augmented with $b$ (similarly, let $C_B = \{C \cup \{a\} | C \subseteq V \wedge b \in C \wedge a \notin C \wedge C \text{ is stable-dense}\}$). Let $C_0 = C_A \cup C_B$; this is the set of all subgraphs that will be examined via cheap-exploration only.

Let $C_{AB} = \{C \cup \{y\} | C \subseteq V \wedge a, b \in C \wedge \text{C is stable-dense} \wedge y \text{ is a neighbor of some node in } C\}$ be the set of graphs consisting of a stable-dense subgraph containing $a$ and $b$, augmented with some other node; this is the set of all subgraphs that will be examined via a single exploration iteration.

Let $C_1 = C_0 \cup C_{AB}$; this is the set of graphs containing $a$ and $b$ that consist of a stable-dense subgraph, augmented with one node.

For $i > 1$, let $C_i = \{C \cup \{y\} | C \in C_{i-1} \wedge \text{C is newly-dense} \wedge y \text{ is a neighbor of some node in } C\}$. $C_i$ is the set of graphs containing a newly-dense subgraph that contains $a, b$, and is discoverable after $i$ exploration iterations.

### 4.1 When is a Single Exploration Sufficient ?

Let us now provide a sufficient condition for all newly-dense subgraphs $C$ of cardinality $|C| = n \geq 3$ to contain a stable-dense subgraph of cardinality $n - 1$. Specifically, it is sufficient that:

$$\delta \leq (n-2)(n-1)(g_n \cdot T_n - g_{n-1} \cdot T_{n-1}) \quad (1)$$

(recall that $g_n = \frac{S_n}{n \cdot (n-1)}$, and that the properties of $T_n$ guarantee that the above bound on $\delta$ is strictly positive)

**Proof sketch:** (pigeonhole argument) If all $n-1$ subgraphs of $C$ were sparse before the update, then the contributions of each vertex in $C$ to $dens^-(C)$ should be large. Hence, $C$ must be very dense. However, $C$ was sparse before the update. Thus, the update must have been very large. If the update is not very large, then there will exist an $n-1$ subgraph that was dense before the update.

**Corollary:** The $n-1$ subgraph of $C$ that was dense before the update will either not contain one of $a$ or $b$ (so augmenting it with that vertex will yield $C$), or it will contain both $a$ and $b$. Consequently, for values of $n$ where Equation 1 holds, all newly-dense subgraphs of cardinality $n$ will be contained in $C_A \cup C_B \cup C_{AB} = C_1$.

### 4.2 Instantiating $T_n$

Based on the form of Equation 1, let us now propose a convenient instantiation for $T_n$, that will satisfy the requisite monotonicity properties, while greatly simplifying the bounds we subsequently derive, thus providing additional intuitions. Specifically, the instantiation of $T_n$ that will be used throughout this work is:

$$T_n = \frac{1}{g_n}\left(g_{N_{max}} \cdot T + \delta_{it} \cdot \left(\frac{n-2}{n-1} - \frac{N_{max}-2}{N_{max}-1}\right)\right) \quad (2)$$

where $\delta_{it}$ is a tunable parameter. Note that this is a reasonable value for $T_n$ from a maintenance perspective; for instance, if $S_n = n$, then $T_n = (n-1)T_2 + (n-2)\delta_{it} = \frac{n-1}{N_{max}-1}(T+\delta_{it}) - \delta_{it} = O(n)$, while if $S_n = n(n-1)$, then $T_n = T_2 + (1-\frac{1}{n-1})\delta_{it} = T - \delta_{it}(\frac{1}{n-1} - \frac{1}{N_{max}-1}) = O(1)$.

Importantly, this instantiation results in a much simplified form of Equation 1, specifically $\delta < \delta_{it}$. In the following, we will leverage this fact, to obtain a bound on the number of exploration iterations that DYNDENS needs to perform.

Moreover, for our proposed techniques to be meaningful, it must be the case that $T_n >> 0 \, \forall n \in \{2, \cdots, N_{max}\}$. This, along with the above simplified form of Equation 1, leads to the following validity range for $\delta_{it}$: $\delta_{it} \in (0, \frac{S_{N_{max}}T}{N_{max}(N_{max}-2)})$. The lower bound would correspond to maintaining the smallest possible number of subgraphs, and the upper bound to maintaining most subgraphs (specifically, all subgraphs of cardinality $N_{max}$, and most subgraphs of lower cardinalities) - realistically speaking, one should not set $\delta_{it}$ to any value close to its upper bound.

### 4.3 Bounding the Number of Iterations

We are now able to extend Equation 1, to cases where $\delta > \delta_{it}$.

Specifically, we will show that all newly-dense subgraphs of cardinality $n$ are contained in $C_0 \cup C_1 \cdots \cup C_{\lceil\frac{\delta}{\delta_{it}}\rceil}$, thus in order to compute all newly-dense subgraphs, it is sufficient to explore around stable-dense and newly-dense subgraphs contained in $C_0 \cup C_1 \cup \cdots \cup C_{\lceil\frac{\delta}{\delta_{it}}\rceil}$.

**Proof sketch:** An update of magnitude $\delta$ is equivalent to $\lceil\frac{\delta}{\delta_{it}}\rceil$ updates of magnitude up to $\delta_{it}$; furthermore, re-exploring stable-dense subgraphs will not yield any new dense subgraphs, thus only newly-dense subgraphs will need to be explored subsequently.

**Discussion:** As witnessed from the above result, the magnitude of $\delta$ is directly correlated with the impact on dense subgraphs. A useful analogy is that of an edge weight update as a perturbation: the greater its magnitude $\delta$, the further away in the graph its effects can be potentially felt (i.e. the further away dense subgraphs will need to be explored).

In this context, parameter $\delta_{it}$ offers a tunable space-time trade-off. By setting it to higher values, more dense subgraphs will be maintained, but fewer exploration iterations will be required per edge update. By setting it to lower values, the space overhead (i.e. the number of dense subgraphs maintained that are not output-dense) can be made minimal: nearly 0 for AVGWEIGHT, and comparable to an offline approach otherwise[8]. Consequently, selecting an optimal good value for $\delta_{it}$ is data-dependent; in practice, we observe that DYNDENS performs well for a wide range of $\delta_{it}$ values.

## 5. EVALUATION

Let us now discuss the experimental validation of our techniques. We will first briefly go over the experimental setup. In Section 5.1 we will present experimental evidence for the feasibility of real-time story identification via ENGAGEMENT, as well as the scalability of our proposed approach. We will also examine the main factors that contribute to the efficiency of DYNDENS.

As we have seen throughout this work, there is a lack of existing techniques for efficiently addressing ENGAGEMENT. Nevertheless, in Section 5.2 we evaluate adaptations of relevant techniques to this problem, so as to have a basis for comparison.

Finally, although efficiency has been our main focus in this work, in Section 5.3 we present some qualitative results that highlight the effectiveness of our approach.

**Experimental setup:** All algorithms evaluated were implemented in Java, and executed on 64-bit Hotspot VM, on a machine with 8 Intel(R) Xeon(R) CPU E5540 cores clocked at 2.53GHz. In our experiments, only one core was used, and the memory usage of the JVM was capped at 25G of RAM (the actual memory consumption was typically lower). Finally, in all performance experiments, the time reported is the median time of 3 identical runs.

**Datasets:** Unless otherwise noted, all our experiments were run using real-world datasets, based on a sample of all tweets for May 1st, 2011 (Our dataset consisted of 13.8M tweets. The sampling was performed by Twitter itself, as part of the restricted access provided to its data stream; for details cf. tinyurl.com/twsam). From these, we removed non-English tweets, and tweets that were labeled as spam (using an in-house tweet spam filter [24]), resulting in 3.8M tweets. Subsequently, we used an in-house entity extractor [3] to identify mentions of real-world entities (such as people, politicians, products, etc). 76.5% of the tweets did not mention any entity of interest; 18.3% mentioned one; 4.3% mentioned two, and under 1% mentioned three or more entities. The entire procedure took under 1h 20' (under 350 $\mu$sec per tweet on average).

**Measuring correlation:** Given these sets of co-occurring entities, there are many ways in which entity association can be measured; our techniques are equally applicable, irrespective of the measure used. For our evaluation, we selected two measures from the literature that we found to yield meaningful results under diverse circumstances: a combination of the $\chi^2$ measure and the correlation coefficient inspired by [5] (weighted dataset), that has been found to be highly effective in identifying stories in the blogosphere, as well as a thresholded variant of the log-likelihood ratio [26] (unweighted dataset) that has been successfully used to identify stories in Grapevine over an extended period of time. In general, we note that any measure that measures strength of pairwise association, based on entity occurrences and pairwise co-occurrences can equally be used by our techniques.

**Identifying emerging stories:** Since the goal of our techniques

---
[8]All exact offline approaches, to the best of our knowledge, utilize some form of a growth property, hence need to compute as many subgraphs as DYNDENS with $\delta_{it} \simeq 0$

is to identify stories in real-time, i.e. "stories happening now", a mechanism for discounting older stories is required. To achieve this, we modify our measures of correlation, by applying exponential decay to all entity occurrences and co-occurrences; for instance, in our experiments we used a mean life for a tweet of 2 hours. Note that our techniques are equally applicable without applying any decay, but the stories identified would then correspond to "cumulative stories to date" (cf. Table 3 showing stories for the entire day) as opposed to "current emerging stories" (cf. online demo www.onthegrapevine.ca/now.jsp).

**Approximating complex association measures:** Finally, for many measures of association (e.g. statistical measures, such as the log-likelihood ratio), the appearance of a document with just a single entity, can influence the weight of all edges in the graph (e.g. the log-likelihood ratio of a pair of entities is a function of the number of documents that have appeared to date). This would pose a significant challenge to incremental computations; to overcome it, we make use of the following approximation, that is applicable to any measure: the weight of an edge connecting entities $e_1, e_2$ is computed by ignoring all documents that have appeared after the latest time that either $e_1$ or $e_2$ appeared in some document.

Intuitively, this will not significantly affect edges connecting popular entities; indeed we observed that in practice the resulting drop in precision entailed by this approximation was fairly low[9]. Importantly, this approximation enables us, after observing a document that mentions entities $e_1, \cdots, e_j$, to only update the weights of edges that are incident to at least one of these entities are updated, i.e. only the weights of edges $\{(e_i, X) | i \in \{1, \cdots, j\}, X \in V\}$ will be updated.

Taking the above into account, the precise manner in which our experimental datasets were created is as follows.

For every tweet where at least one entity was identified, entity occurrences and co-occurrences were updated (taking exponential decay into account, with a mean tweet life of two hours). Thereafter, in the case of the weighted dataset, the $\chi^2$ and correlation coefficient of salient entity pairs was updated; the updated edge weight was computed as $max($ correlation coefficient $, 0)$ if $\chi^2$ showed significant correlation ($p < 5\%$), and 0 otherwise. This procedure resulted in 952K positive and 40.5M negative edge weight updates (recall that the latter are very cheap to process).

In the case of the unweighted dataset, the log-likelihood ratio of salient entity pairs was updated. Two entities were connected with an edge iff each entity appeared in at least 5 tweets, and log-likelihood showed significant correlation ($p < 1\%$). This procedure resulted in 43K positive edge weight updates (edge additions), and 41K negative ones (edge removals).

In either case, this step took under 90 seconds for the entire day. The streams of edge weight updates were loaded to memory before initiating our experiments, and the updates were provided to DYNDENS sequentially, and in-memory. This reflects the expected usage of DYNDENS, as the edge weight updates that constitute its input will typically be generated by another process in real-time. All times reported correspond to the time required to process all edge weight updates resulting from a dataset, while maintaining output-dense subgraphs after each update. Specifically, they do not include the time required to preprocess the dataset (e.g. entity extraction, correlation computation), nor do they include the fixed initialization costs of DYNDENS (such as JVM initialization and initialization of necessary indexing structures). It is worth noting, however, that the throughput of DYNDENS can more than match the stream rate, even after factoring in all preprocessing steps (in total, the overhead for all preprocessing and execution of DYNDENS for our dataset of one day was generally under 90 minutes; moreover the most costly preprocessing steps - i.e. named entity extraction- are inherently parallellizable).

### 5.1 Efficiency and Scalability

Let us now examine some of our experimental findings. Figures 4(a)-4(d) show the time required to process all updates from either dataset, for a variety of definitions of density (experiments involving additional density functions can be found in [4]), and for a wide range of values of density threshold $T$, maximum dense subgraph cardinality $N_{max}$. In these figures, $\delta_{it}$ has been set to 1% of its maximum value, given the values of the other parameters (thus the number of maintained dense subgraphs is typically close to the number of output-dense subgraphs). All runs were capped at 10 minutes (runs that took longer than that were terminated); all figures are cropped to exclude such time-outs[10].

We observe that DYNDENS is able to very efficiently process large datasets, across a wide range of useful operating parameters, validating its applicability for efficiently addressing ENGAGEMENT. The chosen parameters range from instances with none, or only a few output-dense subgraphs, to instances with too many output-dense subgraphs (in the thousands); i.e. the extremal parameter values correspond to instances of less practical interest. Interestingly, one can observe a sharp increase in performance beyond certain values of parameters $T$ and $N_{max}$. This is due to the ensuing sharp drop in the average number of output-dense subgraphs. For instance, with reference to Figure 4(c), the average[11] number of output-dense subgraphs of cardinality at most 6, for $T = 1$ is 3.4K; for $T = 0.8$ it is 13.4K; while for $T = 0.7$ it is over 52K. Similar trends can be observed in the other figures as well; cf. [4].

Having discussed the scalability and efficiency of DYNDENS, let us now turn to evaluating its inner workings. Firstly, let us examine the effects of the $\delta_{it}$ parameter. Recall that, low values of $\delta_{it}$ correspond to DYNDENS materializing fewer dense subgraphs, and, correspondingly, having to perform potentially more explorations. In our experiments, we found our techniques to perform equally well for a wide range of values of $\delta_{it}$; however, selecting a value for it, based on characteristics of the dataset can be beneficial to performance. In Figure 4(e), we show the time taken by DYNDENS to process the unweighted dataset (note the semilog scale), for $N_{max} = 10$ and AVGWEIGHT, across all possible values for $\delta_{it}$ (shown normalized to its maximum value for each threshold). We observe an interesting local optimum wrt. $\delta_{it}$, arising from the tradeoff of having to materialize more subgraphs, while enabling faster updates; i.e. increasing $\delta_{it}$ improves performance, up to a point where the additional dense subgraphs that need to be maintained make this a performance drain. For instance, this point is around 0.2 for $T = 0.8$, around 0.1 for $T = 0.9$, and around 0.6 for $T = 1$. It is also interesting to note that this tradeoff comes into

---

[9]Specifically, we measured the error entailed by this approximation, i.e. the absolute difference of the approximated value of each edge weight, minus the actual value of the correlation measure, for all edges, at 100 uniformly distributed time instants. The median error over all edges was invariably 0; the average absolute error over all edges and all time instants was 0.0003 for the weighted dataset, and 0.002 for the unweighted one, and the average relative error was 10% and 6% respectively.

[10]The only data points that had terminated runs are outside the displayed range; these instances had too large a number of output-dense subgraphs, as a result of unrealistic values for $T, N_{max}$ and/or $\delta_{it}$, and were not expected to finish in a reasonable time

[11]Averaged over all updates, and excluding output-dense subgraphs that are not represented in the index, e.g. most too-dense subgraphs, augmented with a non-neighboring node (cf. Section 3.2.3).

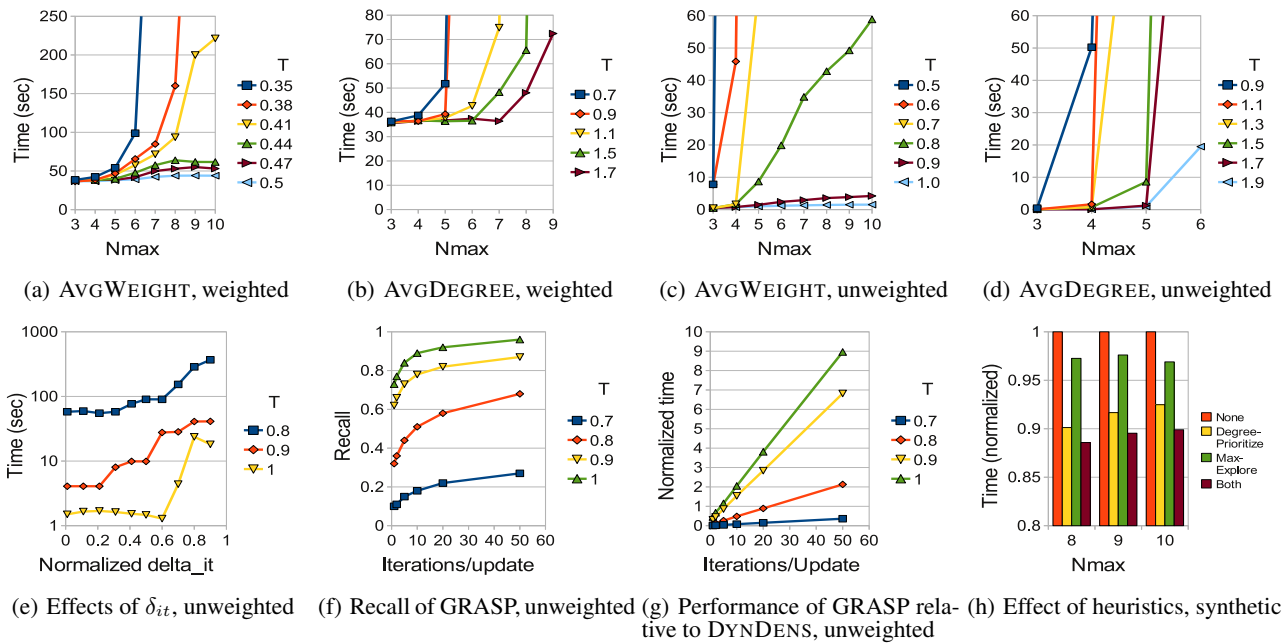

(a) AVGWEIGHT, weighted  (b) AVGDEGREE, weighted  (c) AVGWEIGHT, unweighted  (d) AVGDEGREE, unweighted

(e) Effects of $\delta_{it}$, unweighted  (f) Recall of GRASP, unweighted  (g) Performance of GRASP relative to DYNDENS, unweighted  (h) Effect of heuristics, synthetic

**Figure 4: Experimental evaluation**

play again for $T = 1$ and high $\delta_{it}$.

As we previously saw in Section 3.2.3, IMPLICITTOODENSE is crucially important for DYNDENS to operate efficiently, in the presence of too-dense subgraphs. We validated this intuition experimentally, by executing a variant of DYNDENS that did not make use of IMPLICITTOODENSE, on the weighted dataset, and comparing its runtime to that of DYNDENS. We experimented with execution parameters ($N_{max} \in \{9, 10\}, T \in [0.44, 0.5]$) and with $\delta_{it}$ between 1% and 50% of its maximum value, given the values of the other parameters. Invariably, the variant without IMPLICITTOODENSE took longer than 20 minutes to complete (and was killed after 20 minutes, in the interests of brevity), while DYNDENS took 40-85 seconds to complete.

## 5.2 Comparison with Other Techniques

As we have already discussed throughout this work, to the best of our knowledge, prior to DYNDENS, no techniques have been proposed for efficiently addressing ENGAGEMENT in its general form. Thus, in order to have a basis for comparison, in this section we evaluate adaptations of relevant techniques to subsets of ENGAGEMENT, namely the dynamic maximal clique algorithm proposed in [27] (STIX), the Greedy Randomized Adaptive Search Procedure used to identify large quasi-cliques in [1] (GRASP), as well as a baseline efficient offline procedure that periodically recomputes all AVGWEIGHT dense subgraphs (BASELINE). We wish to stress that, by its very nature, these comparisons are not fair, as the goals of the aforementioned techniques are entirely different from those of ENGAGEMENT, while said techniques are not as general as DYNDENS.

Let us review each comparison in detail. The STIX algorithm [27] identifies all maximal cliques in dynamic unweighted graphs. This is similar to ENGAGEMENT for $T = 1$, AVGWEIGHT and unweighted graphs, but subtly different, in that ENGAGEMENT requires the identification of *all* cliques. Recall that the output of ENGAGEMENT will be used to present stories to a human user, thus the subgraphs produced cannot be too large. If STIX were used to address ENGAGEMENT, and a maximal clique of cardinality e.g. 20 were identified, all its subgraphs of cardinality e.g. 5 or less would need to be enumerated, and provided as output.

Keeping in mind the caveats above, we implemented STIX using an efficient in-memory hash-based index[12], and executed it on the unweighted dataset, measuring its execution time, and ignoring the time that would be needed for enumerating all subgraphs of maximal cliques. We compared this runtime to DYNDENS with AVGWEIGHT, $T = 1$ (so as to have a basis for comparison), $N_{max} = 5$,[13] and set $\delta_{it}$ to half its maximum value, given the values of the other parameters.

Even though a comparison of STIX and DYNDENS is entirely artificial, the runtime of STIX and DYNDENS were roughly equal: STIX took 958 seconds to process the dataset, compared to 936 sec for DYNDENS. DYNDENS performed even better for lower $N_{max}$, and took more time for higher $N_{max}$. Thus, we conclude that DYNDENS is best suited to applications of ENGAGEMENT, while STIX is preferable for applications that require identifying maximal cliques in unweighted subgraphs.

Let us now review the comparison to GRASP, proposed in [1]. This is an approximate randomized algorithm for identifying large dense subgraphs in unweighted graphs. While [1] has significantly more general contributions, for the purposes of this discussion, the algorithm proposed therein can be used to identify subgraphs with density over a given threshold $T$, under AVGWEIGHT, in unweighted graphs. GRASP will not necessarily identify all dense subgraphs, but can be executed multiple times per update, to identify an increasingly larger number of such subgraphs. It is important to note that, again, the comparison with DYNDENS is not straightforward, as GRASP is geared towards identifying a few large dense subgraphs, as opposed to all dense subgraphs.

---

[12][27] does not provide indexing details, so we opted for an efficient solution, albeit with high memory consumption. We also experimented with an adaptation of STIX that used our proposed index, which has much lower memory requirements, but this invariably resulted in increased runtime for STIX.

[13]Since the goal is story identification, we set $N_{max}$ to a low value, corresponding to story cardinalities suitable for humans.

**Table 3: Top stories, May 1st 2011**

| |
|---|
| **Pres. Obama announces killing of Osama bin Laden** *involving:* Barack Obama,U.S. House Permanent Select Committee on Intelligence,Osama bin Laden,NBC News |
| **Commentary on death of bin Laden, comparison to famous athletes** *involving*[14]: Barack Obama,LeBron James,Delonte West,Osama bin Laden |
| **Discussions on Lady Gaga's activities** *involving:* Lady Gaga,Galeria |
| **Libya crisis:NATO Airstrike results in death of 3 grandchildren of Gaddafi** *involving:* NATO,Libya |
| **Discussions on Harry Potter** *involving:* Hermione Granger,Draco Malfoy,Bella Swan |
| **News on Osama Bin Laden's Death Spreads On Twitter** *involving*[15]: Clint Eastwood,Barack Obama,U.S. House Permanent Select Committee on Intelligence,Osama bin Laden,CBS News |

Nevertheless, we implemented GRASP, using an efficient hash-based in-memory index [16]. We set the parameter $\alpha$ that controls its greediness vs. randomness tradeoff to 0.5, after ensuring this did not result in any significant performance differences[17]. We executed GRASP on the unweighted dataset, for a varying number of iterations per edge weight update (more iterations mean higher runtime, and a higher likelihood of identifying more dense subgraphs), and measured its runtime, and recall (fraction of output-dense subgraphs that it identified, excluding disconnected subgraphs, which it does not produce). We limited GRASP to searching for subgraphs of cardinalities up to $N_{max} = 5$, and normalized the runtime of GRASP to the runtime of DYNDENS for the same parameters[18] (i.e. the normalized runtime of DYNDENS is 1). The normalized runtime of GRASP is reported in Figure 4(g), and its recall in Figure 4(f). As we can see, GRASP offers a runtime/recall tradeoff, and can thus be at times more efficient than DYNDENS (however, in such cases, it offers recall of under 80%). Moreover, GRASP offers diminishing returns wrt. recall (i.e. it takes increasingly many iterations to achieve arbitrarily high recall; even though the increase in runtime is linear wrt. the number of iterations, the increase in recall is decidedly sublinear). Thus, in this context, GRASP is best suited to identifying a sample of all dense subgraphs. However, since high recall is of crucial importance in story identification (missing 20% of important stories would not generally be acceptable), DYNDENS is best suited to addressing ENGAGEMENT in this setting.

Finally, we also investigated a simple baseline approach (BASELINE), which periodically recomputes all output-dense subgraphs wrt. AVGWEIGHT. The aim of this comparison was to validate the necessity for incremental computation as opposed to periodic offline recomputation. We implemented BASELINE using an efficient hash-based in-memory index, and executed it on our experimental datasets with varying parameters ($T, N_{max}$), and at varying uniform sampling intervals (i.e. every $X$ tweets). We measured the number of recomputations that BASELINE was able to perform, given the same time as DYNDENS took for the entire dataset

Even given the above restricted problem setting, we observed that BASELINE was generally not up to the task of *realtime* story identification. In our weighted dataset, and for a wide range of parameters, it was able to perform up to 15-30 recomputations in the same time that DYNDENS processed the entire dataset (corresponding to identifying new stories every 48-96 minutes[19]). In the unweighted dataset (which had on average fewer edges, and was thus more amenable to reprocessing from scratch), BASELINE did somewhat better, performing 135-300 recomputations for the parameters we experimented with (corresponding to identifying new stories about every 5-10 minutes). More detailed results can be found in [4]. We conclude that, although periodic recomputation may be an option in limited scenarios (e.g. unweighted graphs, AVGWEIGHT, not very strict realtime requirements), in general the performance benefits of incremental recomputation are needed to support realtime story identification.

## 5.3 Qualitative Results

Whereas the focus of this work is to efficiently identify dense subgraphs in an incremental manner, we also provide evidence of the effectiveness of our approach. Evaluating the quality of our results for realtime story identification is both inherently challenging, due to the lack of a ground truth for what constitutes an important story for a given medium (e.g. a micro-blogging site vs. a news agency), as well as beyond the scope of this work. We will thus present some sample results of utilizing dense subgraphs for story identification. We have also built a live demo for our techniques, which we will briefly discuss, and encourage interested readers to visit so as to view this work in action.

In order to present sample results, we chose to focus on stories at the granularity of a single day (since presenting stories that were heavily discussed at a specific date and time would be hard to process out of context). We used a dataset similar to the "unweighted" one from our performance experiments, with the following two modifications: entity correlations were computed over the entire dataset, as opposed to using exponential decay; and edge weights were retained for pairs of entities with log likelihood of over 5% significance, rather than being thresholded and restricted to $\{0, 1\}$. We computed dense subgraphs of cardinality up to $N_{max} = 5$, using AVGDEGREE to quantify density, so as to favor larger dense subgraphs; for presentation purposes these were subsequently re-ranked in a diversity-aware manner [2] (subgraph overlap was penalized by multiplying subgraph density by $1 - 0.8 \cdot$ ( fraction of story entities covered by previous stories) ).

Table 3 presents the resulting top stories. We observe that discussions on bin Laden's death feature prominently in the list; moreover, given the typical conversation tone on Twitter, distinct discussions involved comparing the presidential announcement to famous athletes[14], and even the rapid propagation of the news on Twitter. Other stories cover the evolving crisis in Libya, as well as lighter, ongoing issues, such as Harry Potter, and Lady Gaga's antics.

For comparative purposes, we also performed the same procedure on a dataset consisting of all blog posts made on major blog hosting platforms during the same day; due to space constraints the results can be found in [4].

Finally, to validate the effectiveness our approach, we have built a live demo of our techniques, in the context of Grapevine [3]. This prototype processes millions of blog posts on a daily basis, and computes important stories in real-time. It consists of a pipeline that processes blog posts as they are crawled, rejecting spam and non-english language posts, extracts named entity mentions, updates the entity graph, and uses DYNDENS to update the set of current dense subgraphs, as in the "unweighted" dataset used in our experiments. It also maintains track of output-dense subgraphs, which are reported to the user upon request. Besides the entities involved in each output-dense subgraph/story, a few links to relevant blog posts are provided, as well as a link back to Grapevine for further exploration of the historical evolution of the story. Interested readers are encouraged to explore this prototype, available at www.onthegrapevine.ca/now.jsp.

---

[14]A Cleveland blogger compared Osama bin Laden to athlete LeBron James; the discussion continued on Twitter, resulting in a sports-related meme around the death of bin Laden.

[15]C.Eastwood was mentioned in conjunction with this story as part of a humorous meme started by comedian Steve Martin on Twitter.

[16]The index used in [1] is optimized for secondary storage, hence not very useful for the purposes of our comparison.

[17]The average (over the values of all other parameters tested) standard deviation of varying $\alpha \in (0, 1)$ was 4%, and the median standard deviation was 1%.

[18]For DYNDENS we selected a reasonable value of $\delta_{it}$, given the values of the rest of the parameters.

[19]As our dataset corresponds to tweets made in one day.

## 6. HEURISTICS

In concluding our exposition of DYNDENS, let us also examine two additional heuristics that can offer modest performance improvements, without affecting the quality of results. Both are related to limiting the number of explorations, and cheap explorations performed. Due to space constraints, the full details for these, and proofs of their correctness, are omitted, and can be found in [4].

MAXEXPLORE: Whereas it serves to prove the correctness of DYNDENS, the previous bound on exploration iterations that need to be performed on a subgraph $C$ is overly pessimistic, as it is based on several worst-case assumptions. To overcome this challenge, we developed MAXEXPLORE, an improvement over the previous bound, that takes the graph neighborhood of the updated edge, as well as the cardinality of the subgraph being explored, into account. As it is a fairly cheap bound to compute, we can expect MAXEXPLORE to lead to performance improvements in the case of dense subgraphs on which multiple exploration iterations would have otherwise been performed.

DEGREEPRIORITIZE: Another challenge in the basic form of DYNDENS discussed so far, is that a single graph might be explored multiple times, by exploration procedures originating from each of its dense subgraphs. To mitigate the adverse effects this can have on performance, we developed DEGREEPRIORITIZE, a way to organize the search space, and thus often avoid performing redundant explorations, inspired by the degree-based criterion proposed in [28]. At a high level, it guarantees that DYNDENS does not need to explore (or cheap-explore) a subgraph with vertices having dense connections to the subgraph. We thus expect DEGREEPRIORITIZE to offer the greatest benefit to performance in cases of dense subgraphs on which redundant, multiple-iteration explorations would have otherwise been performed.

**Evaluation:** In our evaluation of DYNDENS, the above heuristics were enabled. Thus, to evaluate their performance benefits, we also evaluated variants of DYNDENS where either DEGREEPRIORITIZE and/or MAXEXPLORE were disabled, on both our weighted and unweighted datasets. We observed that these heuristics were responsible for very modest performance improvements of up to 4%, and sometimes even resulted in worse performance.

By design, we expect the proposed heuristics to offer performance benefits in cases where many explorations would have otherwise been performed in their absence. To validate this, and further investigate their potential to improve performance, we evaluated them on a synthetic dataset that consisted of near-cliques, mixed with random edges, that was generated as follows: In an initially empty graph with 100K vertices, 250K updates were generated, each of magnitude $(0, 0.1]$ (with probability 0.3 the update was negative). With probability 0.9, the update occurred within one of 100 predefined sets of 10 vertices each; otherwise, it was uniformly randomly distributed to the remainder of the graph. Finally, in order to evaluate the proposed heuristics in the absence of too-dense subgraphs, updates that would result in too-dense subgraphs for $T = 0.7$ and $\delta_{it}$ at 40% of its maximum value, were rejected.

Figure 4(h) shows the time taken by each DYNDENS variant (no heuristics enabled, only DEGREEPRIORITIZE enable, only MAXEXPLORE enabled, both heuristics enabled), normalized by the time taken by the first variant; the operating parameters were $T = 0.7$, $N_{max} \in \{8, 9, 10\}$, and $\delta_{it}$ at 40% of its maximum value (note that the Y axis does not start at 0). The proposed heuristics are seen to offer performance improvements of up to over 10%; thus, while not as crucial as IMPLICITTOODENSE to performance, we believe that the low effort required to implement these heuristics make them worthwhile for inclusion in DYNDENS.

## 7. RELATED WORK

While we are not aware of any work that addresses the maintenance of dense subgraphs in weighted graphs, under streaming edge weight updates, for a broad definition of density, there exists a rich literature of works dealing with related problems.

[27] addresses incremental maximal clique maintenance, from a mostly theoretical perspective, and using a growth property. This is very closely related to a special case of ENGAGEMENT (namely, for unweighted graphs, AVGWEIGHT, and $T = 1$). An important difference is that our instantiation of ENGAGEMENT deals with *all* cliques, with cardinality constraints, as opposed to maximal cliques of unconstrained cardinality. As discussed in Section 5.2, while the former is better suited to real-time story identification, the latter may be preferable in other scenarios.

[28] addresses near-clique identification, in an offline setting, again from a mostly theoretical perspective, and using a growth property; this corresponds to the offline version of ENGAGEMENT for unweighted graphs, and AVGWEIGHT. The techniques proposed therein cannot be efficiently dynamized in a straight-forward fashion, as the information they rely upon cannot be efficiently maintained across updates. Our DEGREEPRIORITIZE pruning condition is inspired by the parent degree-based criterion proposed in this work. [23] addresses the same problem, using a similar growth property, and with a focus on a parallel implementation. As with the other works, the techniques developed therein are not straight-forward to efficiently dynamize.

**Max (quasi-) clique:** Related problems occur in the maximum clique [25] and quasi-clique literature. To overcome the intractability and inapproximability of this problem, heuristics (typically randomized) have been used to discover large (quasi-) cliques. A crucial difference is that ENGAGEMENT requires the enumeration of *all* dense subgraphs (as from an application perspective, each subgraph corresponds to a story of interest). In contrast, works in the maximum (quasi-) clique domain are geared towards identifying one "good" subgraph per execution iteration. Moreover, most such heuristic techniques are not straightforward to efficiently dynamize.

Perhaps most closely related is the state-of-the-art Greedy Randomized Adaptive Search Procedure used in [1] to identify large dense subgraphs (quasi-cliques). Although this work is more focused towards developing techniques for limited main-memory scenarios, their techniques can be dynamized in an efficient manner to address ENGAGEMENT for unweighted graphs and AVGWEIGHT (cf. Section 5.2).

**Local density:** Other works have dealt with edge-weight update semantics, albeit with much simpler definitions of density. For instance [30] and others maintain dense subgraphs over sliding windows using *neighbor-based patterns* (i.e. whether a dense subgraph should be augmented with an additional node is decided based on local information only). As the problem being addressed therein is very different from ENGAGEMENT, the proposed techniques are inapplicable in the latter domain.

**Max-flow:** [12], [20] and others use (primarily) max-flow based algorithms to identify dense subgraphs. While max-flow algorithms can be dynamized [22], [18], these algorithms can only identify and maintain clusters containing user-specified nodes. In a related vein, [14] uses max-flow to find the top-1 dense subgraph (for AVGDE-GREE); however their techniques cannot be efficiently applied to a top-$k$ or threshold variant, nor can they be efficiently dynamized.

**Dynamic graphs:** Other works (e.g. [10], [6]) have dealt with dynamic graph algorithms under edge weight updates, but do not deal with density problems, focusing instead on properties such as planarity, connectivity, triangle counting, etc. A notable exception is [17], which discusses approximation algorithms to general maximization problems in dynamic graphs. It is, however, theoretical in nature, and its focus is on the approximation ratio of the resulting algorithm, not on efficiency.

**Clustering:** Related problems are also dealt with in the incremental clustering literature (e.g. [11], [15], [8]); however, these deal with graph node insertion and deletion, and the proposed techniques cannot directly accommodate streaming edge weight updates. A tangentially related problem is evolutionary clustering ([7], [21]) which identifies clusters based on both density, and historical data; the goal is to introduce temporal smoothing, so that clusters behave in a stable fashion over time.

**Communities of interest:** [9], and its extension [19], address the problem of supporting efficient retrieval of important 2-neighbors of any node, where the importance of a neighbor is related to local and global edge thresholds. The focus is on better representation of actual interactions, and removal of spurious information, and the provided insights are invaluable for any applications that involve dynamic graphs. However, the problem examined in these works, is substantially different from ENGAGEMENT, hence techniques proposed in these works do not apply in ENGAGEMENT.

**Shingling:** [13] proposes techniques to identify large dense subgraphs in an offline fashion via *recursive shingling*. While this could potentially be dynamized, it is geared towards large subgraphs (100-10K nodes), and would not be effective on smaller subgraphs. [29] also uses LSH to identify cliques of moderate size in large graphs; it is however not easily amenable to dynamization, as it has a significant preprocessing phase.

**Data structures:** Finally, the index structure used by DYNDENS resembles the FP-tree [16], in that both store overlapping subsets in a prefix tree, with inverted lists embedded into the tree structure. However, the FP-tree is optimized for static data, and assumes that tree nodes can be statically ordered in a way that heuristically decreases tree size; this makes it unsuitable for ENGAGEMENT, where tree nodes dynamically change. Moreover, other improvements of the FP-tree over a plain prefix tree are not applicable to ENGAGEMENT, as the problems solved are different.

## 8. CONCLUSIONS

Motivated by the need to mine important stories and events from the social media collective, as they emerge, in this work we examine the problem of maintaining dense subgraphs under streaming edge weight updates. For a broad definition of graph density, we propose the first efficient algorithm, DYNDENS, which is based on novel theoretical results regarding the magnitude of change that a single edge weight update can have. DYNDENS is highly efficient, and able to gracefully scale to rapidly evolving datasets, and we validate the efficiency and effectiveness of our approach via a thorough evaluation on real and synthetic datasets.

Moreover, there are many exciting new directions stemming from this work. For example, an important problem in the social media space is the timely identification of online communities. While it is easy to see how ENGAGEMENT can be applied to this domain, its characteristics are somewhat different from those of real-time story identification (e.g. social graphs are frequently directed, communities are typically subgraphs of larger cardinality than stories, etc.), and it would be interesting to explore how to adapt DYNDENS to the diverse challenges this domain imposes. Another interesting technical problem arises when considering the need for adjusting the density threshold $T$, during execution - e.g. in order to adapt to changes in the dataset. We are actively exploring adapting the techniques used in DYNDENS to more efficiently perform this task.